\documentclass[%
superscriptaddress,
preprint,
 amsmath,amssymb,
 aps,
]{revtex4-2}

\usepackage{graphicx}
\usepackage{dcolumn}
\usepackage{bm}
\usepackage{siunitx}

\begin{document}


\title{Quantifying nanoparticle size effect on the photoacoustic generation efficiency}

\author{Arthur Billon}
\affiliation{%
Sorbonne Université, CNRS, Inserm, Laboratoire d’Imagerie Biomédicale, LIB, 75006, Paris, France.
}%
\affiliation{%
Université Paris Cité, CNRS, Inserm, Laboratoire NABI, 75006 Paris, France
}%

\author{Sarah Boumati}
\affiliation{%
Sorbonne Université, CNRS, Inserm, Laboratoire d’Imagerie Biomédicale, LIB, 75006, Paris, France.
}%
\affiliation{
Université Paris-Saclay, CNRS, Institut Galien Paris-Saclay, 91400, Orsay, France.
}%

\author{Lorenzo Mancini}
\affiliation{
Université Paris-Saclay, CNRS, Institut Galien Paris-Saclay, 91400, Orsay, France.
}%

\author{Charaf Eddine Benaddi}
\author{Elise Michel}
\affiliation{
Université Paris-Saclay, ENS Paris-Saclay, CNRS, PPSM, Gif-sur-Yvette, France.
}%

\author{Clément Linger}
\affiliation{%
Sorbonne Université, CNRS, Inserm, Laboratoire d’Imagerie Biomédicale, LIB, 75006, Paris, France.
}%
\affiliation{
Université Paris-Saclay, CNRS, Institut Galien Paris-Saclay, 91400, Orsay, France.
}%

\author{Colin Schotté}
\affiliation{%
Sorbonne Université, CNRS, Inserm, Laboratoire d’Imagerie Biomédicale, LIB, 75006, Paris, France.
}%

\author{Rachel Méallet}
\affiliation{%
Université Paris-Saclay, CNRS, ISMO, Orsay, France.
}%

\author{Frédéric Gobeaux}
\affiliation{%
Université Paris-Saclay, CEA Saclay, CNRS, NIMBE, UMR 3685, LIONS, Gif-Sur-Yvette, France.
}%

\author{Gilles Clavier}
\affiliation{
Université Paris-Saclay, ENS Paris-Saclay, CNRS, PPSM, Gif-sur-Yvette, France.
}%

\author{Nicolas Tsapis}
\affiliation{
Université Paris-Saclay, CNRS, Institut Galien Paris-Saclay, 91400, Orsay, France.
}%

\author{Jérôme Gateau}
\email{jerome.gateau@cnrs.fr}
\affiliation{%
Sorbonne Université, CNRS, Inserm, Laboratoire d’Imagerie Biomédicale, LIB, 75006, Paris, France.
}%

\date{\today}

\begin{abstract}
Photoacoustic (PA) signal generation in colloidal suspensions of optically absorbing nanoparticles is dominated by the thermal expansion of water for gold nanoparticles, but remains mostly unexplored for organic nanoparticles. Here, we derive a model where the PA generation efficiency scales with particle size and thermoelastic contrast with water. The model is validated using solid lipid nanoparticles labeled with several BODIPY dyes. This experimental validation paves the way for quantitative PA characterization of nanomaterials and rational design of PA contrast agents.
\end{abstract}

\maketitle

Optically absorbing nanoparticles have been investigated in biomedical photoacoustic (PA) imaging~\cite{weber_contrast_2016} since the early 2000s~\cite{10.1117/12.446693}. 
PA nanoparticles (NPs) have been widely developed in different shapes, sizes and core materials -- e.g. carbon~\cite{zhang_carbon_2015}, gold~\cite{mantri_engineering_2020}, polymers~\cite{li_polymer-encapsulated_2014} and, recently, solid lipids~\cite{linger_quantitative_2025,linger_modulation_2025}. However, PA signal generation in colloidal suspensions -- in particular, the light-to-photoacoustic conversion efficiency and its NP size dependency -- remains poorly studied, both theoretically and experimentally, beyond gold nanoparticles (GNPs)~\cite{chen_environmentdependent_2012,shinto_acoustic_2013,prost_photoacoustic_2015,pang_photoacoustic_2016,shi_thermally_2016,mantri_engineering_2020}. 

However, GNPs exhibit special PA features. For aqueous GNP suspensions excited by nanosecond laser pulses, several models have shown that over 90\% of the PA signal is generated in water rather than from the GNPs themselves~\cite{chen_environmentdependent_2012,shi_thermally_2016,prost_photoacoustic_2015}, because of the one order-of-magnitude smaller thermal expansion coefficient of gold at an equilibrium temperature $T_{eq} \gtrsim \qty{20}{\degreeCelsius}$.
Moreover, the energy deposition in the surrounding water is noticeably influenced by the homogeneous temperature inside GNPs.
Additionally, the optical absorption and scattering properties of GNPs are shape and size dependent, complicating the experimental evaluation of the thermal expansion conversion efficiency ($TECE$)~\cite{pang_photoacoustic_2016}. Experimental work with spherical GNPs of different radii $a \le \qty{50}{\nm}$ showed that the ratio between the PA amplitude and $\mu_a$ the absorption coefficient of the suspensions is constant for a low-power optical excitation of duration $\tau_L = \qty{5}{\ns}$~\cite{pang_photoacoustic_2016}, confirming a $TECE$ dominated by water in this operational range.

Modeling and measuring the PA conversion efficiency of colloidal suspensions beyond the specific case of GNPs could benefit quantitative PA approaches~\cite{cox_quantitative_2012} and PA NP design. For organic NPs~\cite{jiang_advanced_2017}, optical and thermoelastic properties are mostly independent (determined by encapsulated chromophores and the core material, respectively), enabling quantitative evaluation of the NP size effect on the PA efficiency -- a gap this work addresses.

The pressure wave generation by PA NPs illuminated with a nanosecond pulse can be divided into two steps with distinct conversion efficiencies. Without loss of generality, we consider aqueous suspensions with negligible water absorption.
First, the incident optical intensity is absorbed and transformed into prompt heat flow at the absorber location with the photothermal conversion efficiency ($PTCE \in [0;1]$). $PTCE$ competes with other deexcitation pathways such as luminescence or photo-chemical reactions. 
Second, heat is converted into transient pressure increase, radiated as acoustic waves. The linear thermoelastic regime, ordinary used in biomedical applications, is assumed here and implies that the PA amplitude is linear with the optical fluence and that no phase transition occurs. 
However, heat may diffuse inside each NP and in the surrounding water during the optical excitation and the acoustic generation. For a solution of molecular absorbers in the stress confinement regime (isochoric conversion), $TECE$ is given by $\Gamma_{wat}$ the Grüneisen parameter of water~\cite{xu_photoacoustic_2006}. This is also true for
NPs with a thermal relaxation time much smaller than the duration of the optical pulse $\tau_L$, because of their small size for instance.
However, for larger particles, $TECE$ may depend on the thermoelastic properties of the NPs and water, and also on the parameters governing the heat diffusion during $\tau_L$. 
Therefore, a correction factor ($\alpha_{_{TE}} \ge 0$) should be applied: $TECE= \alpha_{_{TE}} \cdot \Gamma_{wat}$. 

The PA generation capacity of a colloidal suspension excited with pulsed light can be measured by $\theta^{PA}(\lambda)$ the PA coefficient at the optical wavelength $\lambda$~\cite{lucas_calibrated_2022}:
\begin{subequations}
\begin{eqnarray}
\theta^{PA}(\lambda) = \frac{p_0(\lambda)}{\Phi(\lambda) \cdot \Gamma_{wat}}= PGE \cdot \mu_{a}(\lambda)
\end{eqnarray}
with $p_0$ the initial pressure increase in an optically thin sample, $\Phi$ the optical fluence and $\mu_{a}$ the absorption coefficient. $PGE$ is the adimensional photoacoustic generation efficiency.
\begin{eqnarray}
PGE = PTCE(\lambda)\cdot \frac{TECE}{\Gamma_{wat}} =PTCE(\lambda)\cdot \alpha_{_{TE}}
\end{eqnarray}
\end{subequations}

For monodisperse spherical NPs of radius $a$ excited in the long pulse regime~\cite{prost_photoacoustic_2015,diebold_photoacoustic_1991,khan_photoacoustic_1995,calasso_photoacoustic_2001} and diluted so that there is no overlap between nanoscale heated volumes, $\alpha_{_{TE}}$ can be expressed as~\cite{shinto_acoustic_2013}\footnotemark[1]:
\footnotetext[1]{Supplementary information}
\begin{subequations}
\label{eq:Modelmain}
\begin{eqnarray}
\alpha_{TE} = 1 + \gamma \cdot \chi(a,\tau_L) \textrm{ with } \gamma = \frac{\xi_{_{NP}}}{\xi_{wat}}-1
\end{eqnarray}
where $\chi$ is the heat confinement ratio of a NP ($\chi \in [0;1]$): the energy retained in a NP over the total energy converted into heat by the NP, evaluated at the end of a laser pulse. 
We introduce $\gamma$ the thermoelastic contrast ratio of the suspension, which quantifies the relative difference in $\xi$, named here specific expansion coefficient, between NPs and water.
\begin{eqnarray}
\xi_{i}= \frac{\beta_i}{\rho_i \cdot C_{p,i}} = \frac{\Gamma_{i}}{K_{T,i}} 
\end{eqnarray}
with  $\beta_i$ the thermal expansion coefficient, $\rho_i$ the volumetric mass, $C_{p,i}$ the specific heat capacity at constant pressure, $\Gamma_{i}$ the Grüneisen coefficient and  $K_{T,i}$ the isothermal bulk modulus; $i= \{ wat~\textrm{or}~{NP}\}$ for water and the nanoparticle, respectively.
\end{subequations}

Two independent derivations of Equation~\ref{eq:Modelmain}, based on (1) analytical expressions of PA pressure waves generated by a NP and the heated water~\cite{diebold_photoacoustic_1991,khan_photoacoustic_1995,calasso_photoacoustic_2001} and (2) a scale analysis of the temperature and pressure fields~\footnotemark[1], allowed to validate the previously reported analytical expression~\cite{shinto_acoustic_2013} and also to clarify and specify its operational conditions.
Briefly, NPs are expected to be spherical (or statistically isotropic), monodisperse and made of an homogeneous liquid or isotropic solid. Then, the long pulse regime corresponds to a heat deposition duration
long compared to the transit time of sound across the heated region: $\tau_L \gg a/c$ with $c$ the smallest longitudinal wave velocity between the NP material and water ($c_{wat}$).
Finally, regarding the measurement conditions of $\theta^{PA}$, the sample should be in the stress confinement regime (container size $\gg c_{wat}\cdot\tau_L$) and its particle number density should be large enough for the acoustic field continuity, but low enough so that the NP volume fraction is negligible and the sample is optically thin. Further details are given in supplementary information (\footnotemark[1]~section I). These conditions allow to keep the analytical expression simple. However, the model has not yet been experimentally validated despite several attempts with gold and polymer nanospheres~\cite{aoki_near-infrared_2015,fukasawa_effects_2015}. 

\begin{figure*}
\includegraphics[width=\textwidth]{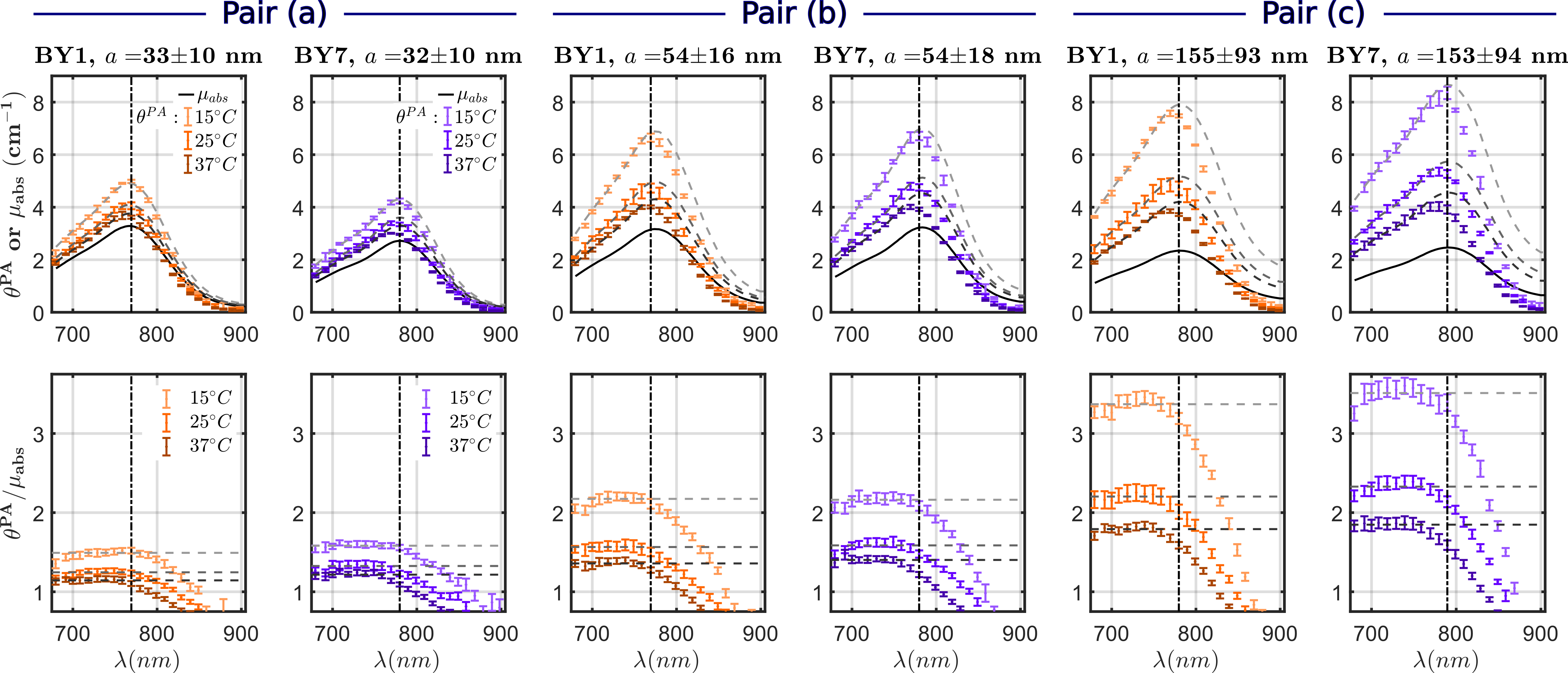}
\caption{\label{fig:theta}
Photoacoustic coefficient $\theta^{PA}$, absorbance coefficient $\mu_{abs}$  and ratio $\theta^{PA}/\mu_{abs}$ of SLN suspensions at three equilibrium temperatures ($T_{eq}=\qty{15}{\degreeCelsius},~\qty{25}{\degreeCelsius},~\qty{37}{\degreeCelsius}$); SLNs formulated with two BY-labels (BY1 and BY7) and in three different sizes ($a \approx$ \qty{33}{\nm} (a), \qty{54}{\nm} (b), \qty{155}{\nm} (c)). All SLNs were prepared at $R_{BY}=10\%$. SNLs labeled with BY1 and BY7 are in shades of orange and violet, respectively. The vertical dashed dotted lines indicate the absorbance maximum wavelength $\lambda_m$. (First row) Absorbance spectrum (black solid line) measured at room temperature and photoacoustic spectrum at three $T_{eq}$. The error bars of $\theta^{PA}$ correspond to the median absolute deviation (MAD) with a scaling factor of 1.4826 over 8 measurements. Dashed lines correspond to $\widehat{PEG_{exp}} \times \mu_{abs}$. (Second row) Ratio $\theta^{PA}/\mu_{abs}$ (ratios below 0.75 are not displayed). The absorbance spectrum is assumed independent of the temperature. The error bars are computed using an uncertainty evaluation~\cite{lucas_quantitative_2023}. The horizontal dashed lines correspond to $\widehat{PGE_{exp}}$, the median value of the ratio over $[\qty{680}{\nm}; \lambda_m]$.}
\end{figure*}

To perform an experimental and quantitative validation, we combine here two key elements: (1) a calibrated PA spectrometer~\cite{lucas_calibrated_2022,lucas_quantitative_2023} designed to obtain quantitative measurements of $\theta^{PA}$ for aqueous suspensions, and (2) colloidal suspensions of solid lipid nanoparticules (SLNs) with a homogeneous spherical inner matrix and formulated in different sizes.

Dexamethasone palmitate (DXP) SLNs labelled with BODIPY dyes were selected because our consortium previously demonstrated that their lipid core contributes to the suspension $TECE$~\cite{linger_quantitative_2025,linger_modulation_2025}.
These particles comprises: a lipid core composed of the lipidic prodrug (DXP) and a BODIPY dye covalently bounded to a palmitate chain (BY-Palm), and an amphiphilic phospholipid-polymer conjugate (DSPE-PEG) located at the particle surface. 
Previous physical studies with these PA SLNs focused on characterizing the spectral and nanostructural transformations that occur when $R_{BY}$ the molar ratio of BY-Palm compared to (DXP+BY-Palm) increases~\cite{linger_quantitative_2025,linger_modulation_2025}. 
Two different BY-Palm labels were investigated: BY-aniline-Palm (BY1), prone to J-aggregation, exhibited strong transformation at $R_{BY} \ge 50\%$, while BY-judolidine-Palm (BY2) was amorphous. 
The present study primarily focuses on the PA characterization of DXP-SLNs with different sizes. Structured aggregation was avoided by setting $R_{BY} \le 25\%$, leading to a spherical, homogeneous and mostly amorphous lipid core~\cite{sarah_preprint_2026,linger_modulation_2025}. 
The SLN radius was modified with two strategies: (1) the formulation of batches by different operators to benefit from inter-operator  particle size dependency, and (2) a reduction of the relative quantity of DSPE-PEG~\cite{lorscheider_nanoscale_2019}. 
To avoid a potential bias linked to the photophysical properties of a specific labelling dye, at least two different BY-Palm were used as SLN labels. As the second strategy failed with BY2 (emulsion demixing), BY7 was introduced as a third BY-Palm~\cite{sarah_preprint_2026}. The low quantum yield of all labels~\cite{sarah_preprint_2026} yields $PTCE \approx 1$. 

A total of 26 samples were formulated and each sample was systematically characterized with three methods~\footnotemark[1]. First, the hydrodynamic diameter was obtained with dynamic light scattering (DLS); more specifically, the intensity weighted Z-average size ($D_Z$) and the polydispersity index ($PdI$)~\cite{iso_DLS_2025}. Second, the absorbance coefficient $\mu_{abs}$ (blanked with water) of a diluted aqueous suspension was measured at room temperature using a UV-Vis spectrophotometer equipped with an integration sphere to minimize scattering signal --- 
$\mu_{abs} \ge \mu_{a}$ with equality when scattering constribution is negligible. Finally, $\theta^{PA}(\lambda)$ of the same suspension was determined with the calibrated PA spectrometer in the range $\lambda \in \left[\qty{680}{\nm} ;\qty{920}{\nm} \right]$ at $T_{eq}=\qty{25}{\degreeCelsius}$. Additionally, 19 samples were measured at $T_{eq}=\qty{37}{\degreeCelsius}$ and 14 at $T_{eq}=\qty{15}{\degreeCelsius}$.

A large discrepancy was previously noticed between $D_Z$ and the mean diameter of the lipid core ($2 \times a$) measured on cryogenic electron microscopy images (cryo-EM) images~\cite{linger_modulation_2025}. An empirical linear relation was determined from measurements performed on 12 samples~\footnotemark[1]:
\begin{equation}
\label{eq:radius}
a \approx \frac{D_Z-\qty{91}{\nm}}{2} \textrm{ and } \frac{\Delta a}{a} \sim \sqrt{PdI}
\end{equation}
With Equation~\ref{eq:radius}, the radius $a$ of the 26 samples spreads over one order of magnitude, from $\qty{15}{\nm}$ to $\qty{250}{\nm}$. Its relative dispersion is between 0.3 and 0.5 for $a < \qty{55}{\nm}$ (19 samples), between 0.5 and 0.6 for $\qty{55}{\nm} < a < \qty{160}{\nm} $ (5 samples) and around 0.7  for $a>\qty{170}{\nm}$ (2 samples). 
The largest nanoparticles, obtained by reducing the amount of DSPE-PEG by a factor 10, are highly polydisperse. 
The mean inter-particular distance was evaluated $\sim \qty{1}{\um}$ for the optically characterized diluted suspensions, satisfying the conditions on separated heated regions around each particle and on the acoustic field continuity for the PA spectrometer~\footnotemark[1].

Figure~\ref{fig:theta} presents $\theta^{PA}$ and $\mu_{abs}$ for six suspensions corresponding to three pairs of SLNs with a similar radius $a$, but prepared with either BY1 or BY7.
The absorbance spectra feature a $S_0-S_1$ electronic transition band with a vibronic shoulder ($S_0-S_1$ vibronic state transition). 
The dye-dependent band maximum wavelength is notated $\lambda_{m}$.
For all suspensions, $\theta^{PA}/\mu_{abs} >1$ and is nearly constant in the wavelength range $\lambda \in [\qty{680}{\nm};\lambda_{m}]$. Above $\lambda_{m}$, the ratio $\theta^{PA}/\mu_{abs}$ decreases with the increasing $\lambda$. This behaviour was previously reported for BY1 and BY2~\cite{linger_modulation_2025}. 
Because $PTCE \approx 1$, the decrease is unexpected given the current PA theory. 
However, a basic analysis of the spectral shapes suggests that, for the smallest SLNs (Figure~\ref{fig:theta}(a-b)), the decreasing ratio could be attributed to a blue shift of a few nanometers between the band maximum in the PA spectrum compared to $\lambda_m$. The shift was previously documented for SLNs with BY1~\cite{linger_quantitative_2025} and a similar blue shift was reported for some fluorescent dyes as Alexa750~\cite{fuenzalida_werner_challenging_2020}, but it is not yet understood. 
For larger SLNs (Figure~\ref{fig:theta}(c)), two additional effects can be noticed: the red portion of the absorbance band centered at $\lambda_m$ is enlarged compared to the PA band and a residual absorbance at $\qty{900}{\nm}$ may indicated an incomplete subtraction of scattering signal with the integration sphere at larger wavelengths.

\begin{figure*}
\includegraphics[width=\textwidth]{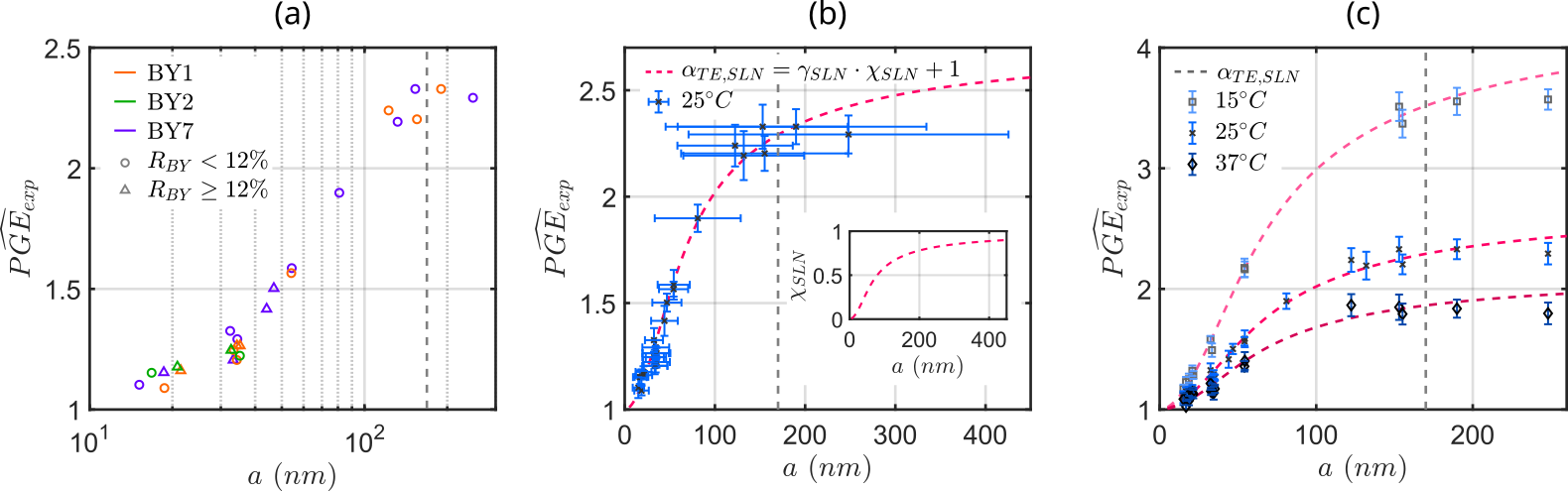}
\caption{\label{fig:PGEexp}
Evolution of $PGE$ with $a$ the SLN radius. (a) $\widehat{PGE_{exp}}$ at $\qty{25}{\degreeCelsius}$ as a function of $a$. The labelling dye is color coded and the molar ratio $R_{BY}$ is coded with two symbols. A logarithmic scale is used for the $a$-axis for a better visual discrimination of markers at low $a$-values. (b) $\widehat{PGE_{exp}}$ as a function of $a$ at $T_{eq}=\qty{25}{\degreeCelsius}$. Vertical error bars correspond to the dispersion of $\widehat{PGE_{exp}}$ evaluated by the maximum value between its calculated uncertainty~\cite{lucas_quantitative_2023} and the median absolute deviation (MAD) of the ratio $\theta^{PA}/\mu_{abs}$ in the range $[\qty{680}{\nm},\lambda_m]$ . Horizontal error bars correspond to Equation~\ref{eq:radius}. The $\chi_{SLN}$-model is displayed in the islet and was used to fit the $\widehat{PGE_{exp}}$ data. (c) $\widehat{PGE_{exp}}$ as a function of $a$ at $T_{eq}= \qty{15}{\degreeCelsius},~\qty{25}{\degreeCelsius},~\qty{37}{\degreeCelsius}$ and fits with the $\chi_{SLN}$-model.    
Vertical gray dashed lines indicate $a=\qty{170}{\nm}$. Fits were performed considering values for $a<\qty{170}{\nm}$.
Some errorbars are omitted in (a) and (c) for the sake of readability.}
\end{figure*}
 
Because of the complexity of the photophysical phenomena inducing a spectral shape mismatch for $\lambda > \lambda_{m}$ and because their investigation is beyond the scope of this paper, a robust estimation of the $PGE$ was obtained by computing the median value of the ratio in the range $[\qty{680}{\nm};~\lambda_{m}]$ and is notated $\widehat{PGE_{exp}}$. 
The use of $\widehat{PGE_{exp}}$ as a $PGE$ estimate is strengthened by the similar values for paired SLNs, at the different $T_{eq}$ (Figure~\ref{fig:theta}).
Figure~\ref{fig:PGEexp}(a) further shows that SLNs of similar radius exhibit consistent $\widehat{PGE_{exp}}$ values at $\qty{25}{\degreeCelsius}$, regardless of the labelling dye and $R_{BY}$ ratio. 
Thereby, only the dependencies of $\widehat{PGE_{exp}} \approx \alpha_{_{TE}}$ with $a$ and $T_{eq}$ are further investigated.

\begin{table}[b]
\caption{\label{tab:eta}%
Estimated thermoelastic parameters of the DXP-SLNs: $\gamma_{SLN}$ and $\xi_{_{SLN}}$, and reference values for palmitic acid and water.}
\begin{ruledtabular}
\begin{tabular}{cccc}
Estimated parameter&
$\qty{15}{\degreeCelsius}$&
$\qty{25}{\degreeCelsius}$&
$\qty{37}{\degreeCelsius}$\\
\colrule
$\xi_{wat}~(\times 10^{-11}\unit{Pa^{-1}})$\footnotemark[1]& 3.6 & 6.2 & 8.7 \\
$\gamma_{palm~acid}$\footnotemark[2] & 3.8 & 1.8 & 1.0\\
\colrule
$\gamma_{SLN}$ & 3.4 & 1.7 & 1.2\\
$\xi_{_{SLN}} (\times 10^{-11} \unit{Pa^{-1}})$\footnotemark[3] & 15.8 & 16.9 & 18.8 \\
\end{tabular}
\end{ruledtabular}
\footnotetext[1]{obtained from tabulated constants of water~\cite{haynes_crc_2017}.}
\footnotetext[2]{obtained from physical constants of palmitic acid at room temperature~\cite{ma14164707,doi:10.1021/j150498a008} and $\xi_{wat}$.}
\footnotetext[3]{computed using $\xi_{_{SLN}}=(\gamma_{SLN}+1) \cdot \xi_{wat}$}
\end{table}

At a given $T_{eq}$, $\widehat{PGE_{exp}}$ significantly increases with the radius $a$ (factor $\approx 2-3$, Figures~\ref{fig:PGEexp}), reflecting the size-dependence of the SLN heat confinement ratio $\chi_{SLN}$. $\chi_{SLN}$ is governed by thermal diffusion and $a$ should then be compared to the thermal diffusion length $\ell_{th}$. In a spherical geometry, $\ell_{th}\sim \sqrt{6 \cdot \kappa \cdot \tau_L}$ with $\kappa$ the thermal diffusivity~\cite{mckenzie_physics_1990}. For $\tau_L =\qty{6}{\ns}$, $\ell_{th} \sim \qty{70}{\nm}$ both in the SLN core and in water, as both media have similar estimated $\kappa$ values~\footnotemark[1]. Thereby, $\ell_{th} \in [\qty{15}{\nm};\qty{250}{\nm}]$ and a large variation of $\chi_{SLN}$ is then expected.
However, the relationship between $a$ and $\chi_{SLN}$ is not linear. 
A semi-analytical thermal diffusion model assuming monodisperse spherical particles~\cite{goldenberg_heat_1952} was implemented to obtain $\chi_{SLN}$ values~\footnotemark[1] and fit experimental data (Figure~\ref{fig:PGEexp}(b)). 
The model considers a homogeneous heating of the SLN during $\tau_L$ and no interfacial thermal resistance, and requires the thermal conductivity and diffusivity of the lipid core and water as input parameters. The thermal properties of palmitic acid~\footnotemark[1] enabled to obtain $\chi_{SLN}$ (islet of Figure~\ref{fig:PGEexp}(b)). 
$\widehat{PGE_{exp}}$ values were fitted according to Equation~\ref{eq:Modelmain} to determine $\gamma_{SLN}$ and $\xi_{SLN}$ (Table~\ref{tab:eta}). 
The two samples with $a>\qty{170}{\nm}$ were discarded due to their high $\Delta a$. 
The model provided an adequate fit at $T_{eq}=\qty{25}{\degreeCelsius}$ (Figure~\ref{fig:PGEexp}(b)), as well as at \qty{15}{\degreeCelsius} and \qty{37}{\degreeCelsius} (Figure~\ref{fig:PGEexp}(c)). Moreover, fitted $\gamma_{SLN}$ values were found similar to $\gamma_{palm~acid}$, the thermoelastic contrast ratio of palmitic acid (Table~\ref{tab:eta}), which validates the PA model. 
Unlike GNPs, where PA signal is dominated by water, SLNs exhibit a significant size-dependent PGE, enabling material-specific tuning.

For a given SLN radius, $\widehat{PGE_{exp}}$ increases with the decreasing temperature. 
This phenomenon is primarily due to the increase of $\xi_{wat}$ with $T_{eq}$ (Table~\ref{tab:eta}). 
The specific expansion coefficient $\xi_{_{SLN}}$ was evaluated at the three $T_{eq}$ and exhibit a slight increase with the temperature (+18\% from $\qty{15}{\degreeCelsius}$ to $\qty{37}{\degreeCelsius}$ compared to +142\% for water).

In sum, from $D_Z$, $\theta^{PA}$ and $\mu_{abs}$ measurements, the lipid core radius $a$ and the $PGE$ were estimated for up to 26 samples labelled with three different dyes. Then, modeling $\chi$ as a function of $a$ enabled to obtain thermoelastic contrast ratios close to that of palmitic acid and estimations of the specific thermoelastic coefficient $\xi_{_{SLN}}$ at three $T_{eq}$. All these steps provide the first experimental demonstration of the forward PA model of Equation~\ref{eq:Modelmain}.

This demonstration further shows that PA sensing is quantitatively sensitive to thermal-expansion phenomena occurring at the nanoscale. 
The influence of the nanoparticle core material, but also its size and the laser pulse duration, on the PA signal amplitude were expressed for homogeneous spherical particles and validated experimentally. 
All these parameters critically influence the light-to-photoacoustic conversion efficiency of colloidal suspensions and must be considered in quantitative PA approaches. However, estimating the heat confinement ratio for new materials remains an open challenge. 
The results initiate quantitative characterization of the thermoelastic properties of nanomaterials with PA spectroscopy and call for further modelling of the PA generation for particles others than GNPs.

\begin{acknowledgments}
This project has received financial support from the French National Research Agency under the programs ANR-21-CE09-0024 and ANR-11-INBS-0006. A. Billon acknowledges funding from the Doctoral School Pierre Louis of PublicHealth.
\end{acknowledgments}

\bibliography{size260707}

\end{document}


\title{Quantifying nanoparticle size effect on the photoacoustic generation efficiency -- Supplementary information}

\author{Arthur Billon}
\affiliation{%
Sorbonne Université, CNRS, Inserm, Laboratoire d’Imagerie Biomédicale, LIB, 75006, Paris, France.
}%
\affiliation{%
Université Paris Cité, CNRS, Inserm, Laboratoire NABI, 75006 Paris, France
}%

\author{Sarah Boumati}
\affiliation{%
Sorbonne Université, CNRS, Inserm, Laboratoire d’Imagerie Biomédicale, LIB, 75006, Paris, France.
}%
\affiliation{
Université Paris-Saclay, CNRS, Institut Galien Paris-Saclay, 91400, Orsay, France.
}%

\author{Lorenzo Mancini}
\affiliation{
Université Paris-Saclay, CNRS, Institut Galien Paris-Saclay, 91400, Orsay, France.
}%

\author{Charaf Eddine Benaddi}
\author{Elise Michel}
\affiliation{
Université Paris-Saclay, ENS Paris-Saclay, CNRS, PPSM, Gif-sur-Yvette, France.
}%

\author{Clément Linger}
\affiliation{%
Sorbonne Université, CNRS, Inserm, Laboratoire d’Imagerie Biomédicale, LIB, 75006, Paris, France.
}%
\affiliation{
Université Paris-Saclay, CNRS, Institut Galien Paris-Saclay, 91400, Orsay, France.
}%

\author{Colin Schotté}
\affiliation{%
Sorbonne Université, CNRS, Inserm, Laboratoire d’Imagerie Biomédicale, LIB, 75006, Paris, France.
}%

\author{Rachel Méallet}
\affiliation{%
Université Paris-Saclay, CNRS, ISMO, Orsay, France.
}%

\author{Frédéric Gobeaux}
\affiliation{%
Université Paris-Saclay, CEA Saclay, CNRS, NIMBE, UMR 3685, LIONS, Gif-Sur-Yvette, France.
}%

\author{Gilles Clavier}
\affiliation{
Université Paris-Saclay, ENS Paris-Saclay, CNRS, PPSM, Gif-sur-Yvette, France.
}%

\author{Nicolas Tsapis}
\affiliation{
Université Paris-Saclay, CNRS, Institut Galien Paris-Saclay, 91400, Orsay, France.
}%

\author{Jérôme Gateau}
\email{jerome.gateau@cnrs.fr}
\affiliation{%
Sorbonne Université, CNRS, Inserm, Laboratoire d’Imagerie Biomédicale, LIB, 75006, Paris, France.
}%

\date{\today}

\maketitle

\tableofcontents

\section{\label{sec:PA_mod}Theoretical photoacoustic model}

The theoretical photoacoustic (PA) model aims at providing an analytical formula for the thermal expansion conversion efficiency ($TECE$) of a colloidal suspension of optically absorbing nanoparticles, given the physical properties of the continuous phase (liquid) and dispersed phase (nanoparticles). 

The PA model is within the linear regime of thermal expansion, which means that no phase change is expected in the absorbing nanoparticles or in the surrounding liquid and that the physical constants at the equilibrium temperature are used in the equations and considered constant over time. This linear regime also implies that the PA generation is considered to be linear with the optical fluence. 
We consider optical excitation of the suspension with nanosecond laser pulses and measurement conditions of the PA spectrometer~\cite{lucas_calibrated_2022} (section~\ref{sec:spectroPA}) developed to obtain $\theta^{PA}$ the PA coefficient  and $PGE$ the PA generation efficiency .

Two independent derivations of the same PA model are presented to strengthen its validation. The first derivation (\ref{sec:Mod1}) is based on analytical solutions of PA pressure waves emitted by spherical nanoparticles and the second derivation (\ref{sec:Mod2}) uses a scale analysis of the spatial distribution of the temperature and pressure fields in the colloidal suspension. The two approaches explicate complementary operational conditions (\ref{sec:Mod3}).

\subsection{\label{sec:Mod1} Derivation from analytical expressions of the PA pressure waves}

The first derivation starts with the analytical expression of the PA pressure wave for a single nanoparticle and then addresses the collection/suspension of nanoparticles.

\subsubsection{Photoacoustic signal from a single spherical particle in the long pulse regime}
A few analytical expressions of the PA pressure wave emitted in a homogeneous liquid by a single optically absorbing spherical particle are available in the literature. These expressions could be obtained in the case of spherically symmetric deposition of heat. Moreover, in the frame of the PA model, we restrict our review to the low frequency approximations corresponding to a transit time of sound across the heated region short compared to the duration of the heat deposition. This approximation is also referred to as the long pulse regime. We performed substantial rearrangements of published formulas to extract common expressions for two cases: the PA pressure wave is generated by the liquid and the PA pressure wave is generated by the nanosphere.

In the context of photoacoustics, the heat deposition arises from optical absorption and is therefore proportional to the fluence rate $\phi(r,t)$ (power per unit surface) where $r$ is the radial coordinate for a spherical coordinate system centered on the sphere and $t$ is the time. To obtain a spherically symmetric deposition of heat, the fluence rate is expected to have a spherically symmetry. The incident pulsed light in the sphere is described by the following expression:
\begin{equation}
 \phi_{sph}(r,t) = \Phi_{sph}(r)\frac{1}{\tau_L}f\left(\frac{t}{\tau_L} \right)
 \label{eq:Phi}
\end{equation}
where $\Phi_{sph}(r)$ is the fluence (energy per unit surface) and $f$ is a dimensionless peaked function describing the temporal profile of the fluence rate. The peak of the function $f$ defines the time $t=0$. $f$ verifies $ \int_{-\infty}^{+\infty} f(\hat{\tau}) \,d\hat{\tau} = 1$ and is normalized by $\tau_L$ which is the full width half maximum and the optical pulse duration. The temporal profile may be modeled as a Gaussian function as:
\begin{equation}
 f(\hat{\tau}) = \frac{2\sqrt{\ln(2)}}{\sqrt{\pi}}e^{-4\ln(2){\hat{\tau}}^2}
 \label{eq:F}
\end{equation}

The case where the PA pressure wave is generated by the heated liquid corresponds to three different scenarios: 1) a ``point absorber'' immersed in the liquid and transferring all the absorbed optical energy converted into heat to the liquid~\cite{calasso_photoacoustic_2001}, 2) an optically absorbing spherical region of the liquid with a spherically symmetric heat deposition~\cite{diebold_photoacoustic_1991}, and 3) an incompressible sphere immersed in a liquid and transferring part of the heat to the liquid which then emits the pressure wave~\cite{diebold_photoacoustic_2002}. The optically thin sphere in the thermal confinement regime providing a homogeneous temperature increase~\cite{diebold_photoacoustic_1988} is a particular case of the second scenario. All three scenarios have a spherically symmetric heat deposition and the emitted pressure wave, at the position $\mathbf{r}$ (coordinate system centered on the sphere) and time $t$, can be expressed as:
\begin{equation}
 p_{liq}(\mathbf{r},t) =   \frac{E_{h,liq} \cdot \beta _{liq}}{C_{p,liq}} \frac{1}{4\pi \lVert {\bf r}\rVert \tau_L }\frac{\partial f}{\partial t} \left(\frac{t - \frac{\lVert {\bf r}\rVert}{c_{liq}}}{\tau_L} \right) 
 \label{eq:Liq}
\end{equation}
where $E_{h,liq}$ is the total heat energy transferred to the liquid (expanding medium) during the pulse duration. $\beta_{liq}$, $C_{p,liq}$ and $c_{liq}$ are the volumetric thermal expansion coefficient, the specific heat capacity at constant pressure and the speed of sound in the liquid at the equilibrium temperature, respectively. 
For the scenario 1), Equation~(\ref{eq:Liq}) can be obtained when 
$\tau_L \gg \kappa_{liq}/c_{liq}^2 $ with $\kappa_{liq}$ the thermal diffusivity in the liquid~\cite{calasso_photoacoustic_2001}, which is also the condition for adiabatic sound propagation~\cite{gousev_laser_1993}. For nanosecond laser pulses, this condition is satisfied in most media including liquids such as water~\cite{gousev_laser_1993}. Indeed, $\kappa_{water}/c_{water}^2 \approx 10^{-13} s$. For scenario 2), the condition corresponds to $\tau_L\gg a/c_{liq}$ with $a$ the radius of the heated spherical region. 
For the scenario 3), both the conditions of $\tau_L\gg a/c_{liq}$ with $a$ the radius of sphere and $\tau_L \gg \kappa_{liq}/c_{liq}^2 $ should be fulfilled~\cite{diebold_photoacoustic_2002}.

In the case where the PA pressure wave is generated inside the sphere and transmitted to the liquid, analytical expressions of the pressure wave in the liquid were obtained for an homogeneous sphere with an homogeneous temperature increase inside the sphere (optically thin and thermal confinement). The analytical expressions were derived for a homogeneous sphere comprised of a second liquid different from the continous phase~\cite{diebold_photoacoustic_1988} and an isotropic solid~\cite{khan_photoacoustic_1995}. The same expression can be extracted for both materials :
\begin{equation}
 p_{sph}({\mathbf r},t) = \frac{\rho_{liq}}{\rho_{sph}} \frac{E_{h,sph} \cdot \beta _{sph}}{C_{p,sph}} \frac{1}{4\pi \lVert {\bf r}\rVert \tau_L } \frac{\partial f}{\partial t} \left(\frac{t - \frac{\lVert {\bf r}\rVert}{c_{liq}}}{\tau_L} \right) 
 \label{eq:Sph}
\end{equation}
where $E_{h,sph}$ is the total heat energy transferred to the sphere during the pulse duration. $\beta _{sph}$, $C_{p,sph}$ and $\rho_{sph}$ are the volumetric thermal expansion coefficient, the specific heat capacity at constant pressure and the volumetric mass inside the sphere at the equilibrium temperature, respectively. $\rho_{liq}$ is the volumetric mass of the surrounding fluid. It can be noticed that Equation~(\ref{eq:Sph}) is independent of the ultrasound wave velocity in the sphere in the long pulse regime. However, Equation~(\ref{eq:Sph}) requires $\tau_L\gg a/c_{sph}$ where $a$ is the radius of the sphere, $c_{sph}$ is the speed of sound inside the sphere for a liquid sphere and the longitudinal wave velocity for a isotropic solid sphere. Comparing Equation~(\ref{eq:Sph}) and (\ref{eq:Liq}), the ratio ${\rho_{liq}}/{\rho_{sph}}$ can be interpreted as a transmission coefficient of the pressure wave from the sphere to the liquid.

Now, we consider the case where the heat deposition inside a homogeneous isotropic solid sphere (or liquid sphere) of radius $a$ has a spherical symmetry, but the heat energy partially diffuses in the surrounding liquid during the optical excitation. The conditions $\tau_L\gg a/c_{sph}$, $\tau_L\gg a/c_{liq}$ and $\tau_L \gg \kappa_{liq}/c_{liq}^2$ are expected to be satisfied. We define $\chi$ the heat confinement ratio of the particle as the ratio between the heat energy retained in the sphere during the pulse duration relative to the total energy converted into heat.
\begin{subequations}
\label{eq:Eh}
\begin{eqnarray}
E_{h,sph} = \chi  \cdot E_{h,tot},
\end{eqnarray}
\begin{eqnarray}
E_{h,liq} = (1-\chi) \cdot E_{h,tot},
\end{eqnarray}
\begin{eqnarray}
E_{h,tot} = PTCE \cdot \Phi_{liq}(\mathbf{r}) \cdot \sigma_a,
\end{eqnarray}
\end{subequations}
where $PTCE$ and $\sigma_a$ are the photothermal conversion efficiency  and  the absorption cross-section of the nanosphere, respectively. $\Phi_{liq}$ is the optical fluence (energy per unit surface) at the scale of the liquid containing the nanoparticle dispersion ($\Phi_{liq}(\mathbf{r}) =\Phi_{sph}(a)$).
Then, the PA pressure wave emitted by the spherical particle and the heated liquid can be expressed as:
\begin{widetext}
\begin{equation}
 p_{sph+liq}({\mathbf r},t) = PTCE \cdot \Phi_{liq} \cdot \sigma_a \cdot \left[\chi\cdot\frac{\rho_{liq}}{\rho_{sph}} \frac{\beta _{sph}}{C_{p,sph}} + (1-\chi)\frac{\beta_{liq}}{C_{p,liq}} \right] \frac{1}{4\pi \lVert {\bf r}\rVert \tau_L } \frac{\partial f}{\partial t} \left(\frac{t - \frac{\lVert {\bf r}\rVert}{c_{liq}}}{\tau_L} \right) 
\end{equation}
\end{widetext}

\subsubsection{Photoacoustic signal from the colloidal suspension of spherical nanoparticles}

 Here, we consider the experimental configuration of the PA spectrometer~\cite{lucas_calibrated_2022}, that is to say that the colloidal suspension is injected in a tube of diameter $D_{tube}$ immersed in water (same liquid as the one surrounding the particles). The colloidal suspension is expected to be homogeneous in the tube and the numeric concentration of nanospheres is notated $n_{sph}$. We assume that $c_{liq} \cdot \tau_L > (n_{sph})^{1/3}$ such that the medium can be considered continuous with regards to the acoustic field. If this condition is not satisfied for ultrasound pulsations on the order of $\omega_{ac} \sim 2\pi/\tau_L$, the pressure wave $p_{sph+liq}({\bf r},t)$ can be convolved in the time domain with the electric impulse response of the ultrasound detector to account for its bandwidth $[\omega_{ac 1};\omega_{ac 2}]$ and obtain $c_{liq} / \omega_{ac 2} > 2\pi(n_{sph})^{1/3}$ without loss of generality. The discrete model is not considered here. The volume fraction of the dispersed phases is expected negligible so that the speed of sound of the suspension is equal to $c_{liq}$.

The pressure wave emitted by the tube containing the colloidal suspension is:
\begin{equation}
 p_{tube}({\mathbf{r'}},t) =  n_{sph} \cdot \iiint_{\bf r \in V_{tube}} p_{sph+liq}({\mathbf{r'}}-\mathbf{r},t) \,d^3\mathbf{r}
 \label{eq:Susp1}
\end{equation}
where $V_{tube}$ is the illuminated portion of the tube.
 
In the tube, the colloidal suspension is considered to be optically thin: $n_{sph} \cdot \sigma_a \cdot D_{tube} \ll 1$, so that the optical fluence $\Phi_{liq}(\mathbf{r})$ is independent of the solution inside the tube. The fluence can however have spatial heterogeneities due to the illumination system.
 
The Grüneisen coefficients of the liquid $\Gamma_{liq}$ and of the sphere $\Gamma_{sph}$ is given by:
\begin{subequations}
 \label{eq:Grun}
\begin{eqnarray}
 \Gamma_{liq} =  \frac{\beta _{liq} \cdot c_{liq}^2}{C_{p,liq}} =\frac{\beta _{liq} \cdot K_{T,liq}}{\rho_{liq} \cdot C_{v,liq}}
 \label{eq:Grun_liq}
\end{eqnarray}
\begin{eqnarray}
 \Gamma_{sph} = \frac{\beta_{sph}\cdot K_{T,sph}}{\rho_{sph}\cdot C_{v,sph}}
 \label{eq:Grun_sph}
\end{eqnarray}
\end{subequations}
where $K_{T,liq}$ and $K_{T,sph}$ are the isothermal bulk modulus of the liquid and the sphere, respectively. $C_{v,liq}$ and $C_{v,sph}$ are the specific heat capacities at constant volume of the liquid and the sphere, respectively. $C_p/C_v \approx 1$ for the liquid and the sphere.

We introduce $\xi$ the specific expansion coefficient :
\begin{equation}
\xi_{i}= \frac{\beta_i}{\rho_i \cdot C_{p,i}} = \frac{\Gamma_{i}}{K_{T,i}} 
\end{equation}
with $i= \{ liq~\textrm{or}~sph\}$ for liquid and the sphere, respectively.

The absorption coefficient corresponding to the spheres in the colloidal suspension is given by:
\begin{equation}
 \mu_a = n_{sph} \cdot \sigma_a
 \label{eq:abs_coef}
\end{equation}
$\mu_a$ can be obtained experimentally with a spectrophotometer equipped with an integration sphere, to be insensitive to scattering.

Then, Equation~(\ref{eq:Susp1}) can be written:
\begin{equation}
 p_{tube}({\mathbf{r'}},t) = \theta^{PA} \cdot \Psi_{tube}({\mathbf{r'}},t)
 \label{eq:Susp2}
\end{equation}
where $\theta^{PA}$ is the PA coefficient~\cite{lucas_calibrated_2022} and $\Psi_{tube}({\mathbf{r'}},t)$ is the acoustic response of the sample container defined in Equation~(\ref{eq:Psi}). This acoustic response is further convolved by the electrical impulse response of the detector and the spatial impulse response to obtain the measured PA signal. The calibration process of the PA spectrometer allows to compute $\theta^{PA}$ from the amplitude of the measured PA signal~\cite{lucas_quantitative_2023}. 

Finally, we obtain the PA model :
\begin{subequations}
\label{eq:PGEtheta}
\begin{eqnarray}
\theta^{PA}=\mu_a \cdot PGE
\label{eq:Theta}
\end{eqnarray}
\begin{eqnarray}
 PGE =  PTCE\cdot \frac{TECE}{\Gamma_{liq}}
 \label{eq:PGE}
\end{eqnarray}
\begin{eqnarray}
 TECE =  \Gamma_{liq}\cdot\left[\chi\cdot \left( \frac{\xi_{sph}}{\xi_{liq}}-1 \right) + 1\right]
 \label{eq:TCEC}
\end{eqnarray}
\end{subequations}
where $PGE$ and $TECE$ are the PA generation efficiency and the thermal expansion coefficient of the suspension, respectively. 

 \begin{widetext}
 \begin{eqnarray}
 \Psi_{tube}({\mathbf{r'}},t) = \frac{\Gamma_{liq}}{c_{liq}^2} \iiint_{\mathbf{r} \in V_{tube}}  \frac{\Phi_{liq}(\mathbf{r})}{4\pi \lVert {\mathbf{r'}-\mathbf{r}}\rVert \tau_L} 
\frac{\partial }{\partial t}f \left(\frac{t - \frac{\lVert {\mathbf{r'}-\mathbf{r}} \rVert}{c_{liq}}}{\tau_L} \right) \,d^3r
\label{eq:Psi}
\end{eqnarray}
 \end{widetext}

\subsection{\label{sec:Mod2}Derivation with scale analysis and consideration of the state equation}
 
At the scale of the sample container, stress confinement can be assumed when $\tau_L \ll D_{tube}/{c_{liq}}$~\cite{xu_photoacoustic_2006}. Then, the PA problem can be written with an initial pressure increase:
\begin{equation}
 p_{tube}({\mathbf{r'},t=0}) =  PTCE \cdot \mu_a \cdot \Phi_{liq}(\mathbf{r'}) \cdot TECE 
 \label{eq:P0}
\end{equation}
Equation~(\ref{eq:P0}) indicates that the spatial scale for the distribution of the initial pressure increase (before propagation of the pressure wave) corresponds to the tube diameter for a uniform illumination or to that of the fluence distribution for heterogeneities linked to the illumination system. The spatial scale of the temperature distribution is much smaller as the temperature increase during the optical pulse is only expected in the nanoparticle and its immediate vicinity. These spatial distributions match the fact that, during the optical excitation, the length for heat diffusion is short compared with the typical propagation length of sound across in the medium (long pulse regime, section \ref{sec:Mod1}).

The pressure field, the density field and the temperature field in the solution are notated: $p$, $\rho$ and $T$, and their equilibrium values: $p_{eq}$, $\rho_{eq}$ and $T_{eq}$, respectively. 
The bulk modulus $K_{T}$ can be expressed as:
\begin{equation}
\frac{1}{K_{T}}=\frac{1}{\rho_{eq}}\left(\frac{\partial \rho}{\partial p}\right)_{T_{eq}}
 \label{eq:KT}
\end{equation}
Because of the spatial distribution of the pressure field $p_{tube}({\mathbf{r'},t=0})$ and when the volume fraction of the nanosphere can be neglected, the effective bulk modulus of the solution can be taken equal to that of the liquid.

The volumetric coefficient of thermal expansion $\beta$ is given by:
\begin{equation}
\beta=-{\frac{1}{\rho_{eq}}\left(\frac{\partial\rho}{\partial T}\right)}_{p_{eq}}
 \label{eq:Beta}
\end{equation}
Because of the spatial distribution of the temperature field, the spatial heterogeneities of $\beta$ should be taken into account.

The expansion of the state equation $p=p(\rho,T)$ near the equilibrium values can be written considering the time variation of the spatio-temporal fields:
 \begin{equation}
\frac{1}{K_{T}}\frac{\partial p}{\partial t}=\frac{1}{\rho_{eq}}\ \frac{\partial\rho}{\partial t}+\beta\frac{\partial T}{\partial t}\ 
 \label{eq:state}
\end{equation}
In the stress confinement regime, the density will not have time to decrease during the laser pulse duration and thereby $\left|\frac{\partial\rho}{\partial t}\right|\ll\left|\frac{\partial T}{\partial t}\right|$. 
Under this isochoric condition, Equation~(\ref{eq:state}) becomes:
 \begin{equation}
\frac{1}{K_{T}}\frac{\partial p}{\partial t} \approx \beta\frac{\partial T}{\partial t}\ 
 \label{eq:state2}
\end{equation}

Additionally, because the thermal confinement is satisfied, the rate at which the temperature increases is given by:
\begin{equation}
\frac{\partial T}{\partial t}=\frac{P_v}{\rho_{eq} C_v}
\label{eq:state3}
\end{equation}
where $P_v$ is the heat power density (power per unit volume).

By integration of Equation~(\ref{eq:state2}) over the duration of the optical pulse and with a spatial averaging over a volume where $\Phi_{liq}$ is homogeneous, we obtain:
\begin{equation}
\frac{p_{tube}(t=0)}{K_{T,liq}} = n_{sph} \cdot \left( 
E_{h,sph} \cdot \xi_{sph} + E_{h,liq} \cdot \xi_{liq}
\right)
\label{eq:state4}
\end{equation}
Combining Equations~(\ref{eq:state4}),~(\ref{eq:P0}),~(\ref{eq:abs_coef}),~(\ref{eq:Grun}) and~(\ref{eq:Eh}), we obtain again Equation~(\ref{eq:TCEC}).

A similar method was used~\cite{shinto_acoustic_2013} to obtain the expression of $TECE$. However, our derivation clarifies the ambiguous assumptions regarding the stress confinement at the “characteristic length of heat heterogeneity”~\cite{shinto_acoustic_2013} and the short relaxation times of the pressure field in the nanosphere and in the surrounding heated medium.

It can be noticed that the spherical symmetry of the optically absorbing particle does not explicitly appear in this demonstration. However, the pressure wave emission arising from the position $\mathbf{r'}$ inside the tube is assumed to be isotropic. Then, it is expected that Equation~(\ref{eq:PGEtheta}) can be applied for particles in Brownian motion with random orientations, regardless of their shape if they are much smaller than the smallest detected acoustic wavelength ($2\pi \cdot c_{liq}/\omega_{ac 2}$).

 \subsection{\label{sec:Mod3} Summary of the operational conditions}
 
 In sum, the conditions for the applicability of the PA model are:
 \begin{itemize}
  \item \emph{Particle material}: liquid or isotropic solid. The model considers a material with an homogeneous specific expansion coefficient $\xi_{sph}$. 
  However, for particles in Brownian motion with random orientations, the homogeneity may be understood in the frame of a statistical isotropy with effective physical constants.
  \item \emph{Particle size}: ideally the characteristic length $a$ should satisfy : $a \ll c \cdot \tau_L$ with $c$ the smallest longitudinal wave velocity between the particle and the surrounding liquid. However, the condition: $a \ll 2\pi \cdot c_{liq}/\omega_{ac 2}$ with $\omega_{ac 2}$ the low pass cutoff pulsation of the detector is expected to be sufficient. 
  The particles are assumed to be monodisperse.
  \item \emph{Particle shape}: ideally spherical, but a statistical isotropy can be assumed when the condition on the particle size is satisfied.
  \item \emph{Particle concentration}: the particle number concentration should be high enough so that the medium is considered continuous with regards to the acoustic field, but low enough so that the volume fraction of the particles is negligible (speeed of sound and density of the solution equal to that of the liquid) and the heated regions during $\tau_L$ do not overlap . 
  \item \emph{Pulse length}: the optical pulse length should satisfy $\tau_L \gg \kappa_{liq}/c_{liq}^2$  and  $\tau_L \ll D_{tube}/{c_{liq}}$.
\end{itemize}

\section{\label{sec:chi_mod}Determination of the heat confinement ratio $\chi$ of a nanoparticle}

This method considers monodisperse homogeneous spheres of radius $a$ embedded in an infinite homogeneous medium. During the laser pulse, we assume that heat is produced uniformly and at a constant rate in the volume of the sphere. This approximation allows to use an semi-analytical solution to obtain the heat energy retained inside the sphere at the end of the optical pulse~\cite{goldenberg_heat_1952}. The solution considers no interfacial thermal (Kapitza) resistance between the particle and the surrounding liquid.

Setting the radial coordinate at the center of a single particle and the time origin at the onset of the laser pulse, the thermal energy inside one particle can be written: 
 
\begin{equation}
\label{eq:ch1}
E_{h,sph}=\rho_{eq,sph}\cdot C_{v,sph}\cdot \int_{r=0}^{a} \Delta T_{sph}(r,\tau_L)\cdot 4\pi r^2 \,dr
\end{equation}
where $r$ is the radial coordinate for a coordinate system centered on the sphere, $\Delta T_{sph}(r,\tau_L)$ is the increase of temperature field with regards to the equilibrium temperature $T_{eq,sph}$ and evaluated at the end of the laser pulse duration. The equilibrium temperatures of the medium and the nanoparticle are taken equal: $T_{eq,sph}=T_{eq,liq}$.

In solids and liquids, $C_p / C_v \sim 1$. 
Moreover, the thermal diffusivity $\kappa$ and thermal conductivity $\Lambda$ are related by:
\begin{equation}
\kappa=\frac{\Lambda}{\rho_{eq}\cdot C_{p,eq}}
\end{equation}

We model the laser pulse as rectangular and evaluate the temperature increase at time $t=\tau_L$ using the formula:
\begin{equation}
\Delta T_{sph}(r,\tau_L) = \frac{a^2 \cdot A}{\Lambda_{sph}}\varphi(r,\tau_L)
\end{equation}
where A is the constant heat rate per unit time and per unit volume and $\varphi(r,\tau_L)$ is defined by Equation~(\ref{eq:varphi}) obtained from reference~\cite{goldenberg_heat_1952}.

For a rectangular pulse, $A$ can be expressed as:
\begin{equation}
\label{eq:ch2}
A = \frac{3}{4\pi a^3\cdot \tau_L}E_{h,tot}
\end{equation}
Thereby, combining Equations~(\ref{eq:ch1}) to~(\ref{eq:ch2}), the heat confinement ratio can be written:
\begin{equation}
\chi(a,\tau_L) = \frac{3}{a\cdot\tau_L\cdot \kappa_{sph}}\cdot\int_{r=0}^{a}{\varphi(r,\tau_L) \cdot r^2 \,dr}	
\end{equation}
 
\begin{widetext}
\begin{subequations}
\label{eq:varphi}
\begin{eqnarray}
 \varphi\left(0<r \le a,\tau_L\right)= \frac{\Lambda_{sph}}{3\Lambda_{liq}}+\frac{1}{6}\left(1-\frac{r^2}{a^2}\right) 
 -\frac{2 a \vartheta_1}{\pi r}
\int_{0}^{+\infty}{\frac{e^{-y^2\cdot\tau_L{\cdot\kappa}_{sph}{/a}^2}\cdot \left(sin(y)-y \cdot cos(y)\right)\cdot sin(r\cdot y/a)}{y^2\cdot\left[\left(\vartheta_2\cdot sin(y)-y\cdot cos(y)\right)^2+\left(\vartheta_1\cdot y\cdot sin(y)\right)^2\right]} \,dy}
\end{eqnarray}
\begin{eqnarray}
\varphi(r=0,\tau_L)=\frac{\Lambda_{sph}}{3\Lambda_{liq}}+\frac{1}{6}-\frac{2\cdot\vartheta_1}{\pi}\int_{0}^{+\infty}{\frac{e^{-y^2\cdot\tau_L{\cdot\kappa}_{sph}{/a}^2}}{y}\frac{\left(sin(y)-y\cdot cos(y)\right)}{\left(\vartheta_2\cdot sin(y)-y\cdot cos(y)\right)^2+\left(\vartheta_1\cdot y\cdot sin(y)\right)^2}\,dy}
\end{eqnarray}
\begin{eqnarray}
\vartheta_1=\frac{\Lambda_{liq}}{\Lambda_{sph}}\cdot\sqrt{\frac{\kappa_{sph}}{\kappa_{liq}}}
\end{eqnarray}
\begin{eqnarray}
\vartheta_2=1-\frac{\Lambda_{liq}}{\Lambda_{sph}}.
\end{eqnarray}
\end{subequations}
 \end{widetext}

The field $\varphi(r,\tau_L)$ was evaluated using a numerical integration with a N-point Gaussian quadrature rule. After analysis of the integrands, we chose to set the upper limit to $10\pi$ and to use $N= 200$. Then, the integration of Equation~(\ref{eq:ch1}) was performed using a rectangle rule with steps of $\qty{0.1}{\nm}$. Numerical simulation was performed in Python.

\section{Experimental section}

\subsection{Material and methods}

\subsubsection{Formulation of the solid lipid nanoparticles}

Solid lipid nanoparticles (SLNs) were formulated by emulsion-evaporation process using a protocol  previously detailed~\cite{linger_modulation_2025,linger_quantitative_2025,sarah_preprint_2026}.
The BY-Palm compounds (BY1, BY2 and BY7) and their synthesis were presented in ref~\cite{sarah_preprint_2026}.

Five different batches of SLNs were formulated by four different operators. Samples were spread over $R_{BY} \in [6\%;25\%]$. The standard experimental protocol was used for 19 samples. The quantity of DSPE-PEG$_{2000}$ was divided by a factor 5 to 15 to increase the SLN size for 7 samples (distributed in 3 batches).

The mother solutions were diluted before optical characterization with UV-Vis spectroscopy and PA spectrometer, to obtain a maximum absorption coefficient $\mu_{abs} \lesssim 3$ \unit{\per\cm}. 

\subsubsection{Dynamic light scattering}
The hydrodynamic size was determined with a Zetasizer Nano ZS (Malvern Instrument, UK) using the intensity weighted Z-average size ($D_Z$) and the polydispersity index ($PdI$)~\cite{iso_DLS_2025}. 
The SLNs were found to have a spherical shape in cryo-EM images~\cite{linger_quantitative_2025,sarah_preprint_2026}. Therefore, the mean hydrodynamic diameter was set equal to $D_Z$. A single size population following a Gaussian distribution was assumed, so that the polydispersity index was transformed into a standard deviation estimate:
\begin{equation}
\label{deltaD_z}
\Delta D_Z = \sqrt{PdI} \cdot D_Z
\end{equation}

\subsubsection{Cryo-EM images}

Cryogenic Transmission Electron Microscopy (cryo-TEM) images were performed according to the protocol reported in~\cite{linger_modulation_2025}. 
The diameters of several hundreds of nanoparticles were measured. The average diameter $D_C$ and the standard deviation $\Delta D_C$ were computed on the $N_C$ measurements (Table~\ref{tab:table3} and~\cite{linger_modulation_2025}).

\subsubsection{Absorbance spectra.}

Absorbance spectra of diluted suspensions were measured using a UV-visible spectrophotometer UV-2600 (Shimadzu, Japan) with an integrating sphere coated with barium sulphate ISR-2600PLUS to minimize scattering signal. Absorbance measurements were performed in transmittance using $L_{cuv} = \qty{10}{\mm}$ pathlength quartz cuvettes. A blank measurement was performed with Milli-Q water.

The absorbance coefficient $\mu_{abs}$ at the optical wavelength $\lambda$ was computed from $Abs(\lambda)$ the absorbance output of the spectrophotometer using:
\begin{equation}
\mu_{abs}(\lambda) = Abs(\lambda) \cdot \ln(10) / L_{cuv} 
\end{equation}

\subsubsection{\label{sec:spectroPA} Calibrated PA spectrometer.}

PA spectra of the diluted SLN suspensions were acquired using a calibrated PA spectrometer, previously described in details~\cite{lucas_calibrated_2022,lucas_quantitative_2023,Davenet:26}. Parameters important for the PA model are given here. 

PA coefficients $\theta^{PA}(\lambda)$ were obtained from $\lambda=\qty{680}{\nm}$ to $\lambda=\qty{920}{\nm}$ in \qty{10}{\nm}-steps. Optical excitation was provided by a tunable nanosecond laser (SpitLight 600 OPO, Innolas, Germany) delivering pulses at a repetition rate of 20 Hz with a pulse duration of 6-7 \unit{\ns}. The pulse duration is set to $\tau_L =\qty{6}{\ns}$ for the models. 

The samples were loaded into PTFE tubes of inner diameter $D_{tube} = \qty{0.2}{\nm}$ (Bola, Germany). Because $\mu_{abs} \cdot D_{tube} \lesssim 6 \times 10^{-2}$, the samples are optically thin. Ultrasound (US) detection was performed between \qty{2}{MHz} and \qty{6}{MHz} with a linear US array (L7-4, ATL, USA)~\cite{lucas_calibrated_2022}. The sample are in the stress confinement regime because $\tau_L \ll D_{tube}/c$ with $c \approx 1.5 \times 10^{3}~\unit{m.s^{-1}}$ the speed of sound in the suspension. The tubes, the tested samples and the ultrasound probe were immersed in a temperature-controlled water bath (T100-ST12 Optima, Grant, UK) for thermalisation at $T_{eq}=\qty{15}{\degreeCelsius},~\qty{25}{\degreeCelsius}$ and \qty{37}{\degreeCelsius}, respectively.

The incident optical fluence on the tubes did not exceed 3.5 \unit{mJ.cm^{-2}}. SLNs loaded with BY1 and BY2 were previously reported to be in the linear thermoelastic regime for this illumination conditions~\cite{linger_modulation_2025}, because similar $\theta^{PA}$ values were reported at maximal fluence of 3.5 and 2.5 \unit{mJ.cm^{-2}}. 

Two calibration solutions were employed: a 0.25M CuSO$_4\cdot$5H$_2$O solution (Sigma-Aldrich, USA) and a naphthol green B solution (Thermo Fisher Scientific, USA)~\cite{Davenet:26}. Additionally, two previously reported validation solutions were used. A solution of nigrosin (Sigma-Aldrich, USA) at a concentration of 173 \unit{mg.L^{-1}}~\cite{lucas_calibrated_2022}. A suspension of gold nanoparticles with a \qty{8}{nm} diameter (Au@DTDTPA) synthesized at Université de Franche-Comté (Besançon, France)~\cite{lucas_quantitative_2023}. The median of the ratio $\theta^{PA}/\mu_{abs}$ in the range $[\qty{680}{\nm};\qty{920}{\nm}]$ was found equal to 1.0 at the three equilibrium temperatures for these validation solutions. 

\subsection{Supplementary experimental results.}

\subsubsection{Hydrodynamic diameter vs cryo-EM diameter.}

The mean diameter of 12 samples were both measured with DLS and cryo-EM. Measurements previously reported~\cite{linger_modulation_2025}
for unlabelled SLNs and for combination Dye-$R_{BY}$ equal to BY1-2\%, BY1-25\%, BY2-2\% and BY2-25\% were used. Table~\ref{tab:table3} displays the values for the additional datasets.  

\begin{figure}[t]
\includegraphics[width=\linewidth]{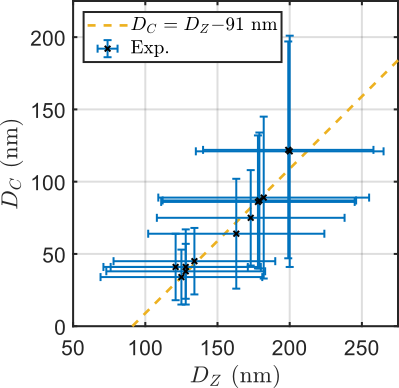}
\caption{\label{fig:DLS_cryo} Mean hydrodynamic diameter $D_Z$ evaluated with dynamic light scattering as a function of the mean diameter $D_C$ evaluated on cryo-EM images for 12 samples. The error bars are estimates of the dispersion of the nanoparticle diameters. A linear fit is plotted with a dashed line.}
\end{figure}

The average diameters exhibit a linear dependency (Figure~\ref{fig:DLS_cryo}). The slope, the intercept, and their standard errors were obtained with the Matlab function \emph{fitlm}. With observation weights inversely proportional to the relative error for $D_C$, we determined: 
\begin{equation}
D_C = \alpha_1 \cdot D_Z +\alpha_2
\end{equation}
with $\alpha_1 = 1.01 \pm 0.07$ and $\alpha_2 =-91\pm$\qty{12}{\nm}.

Consequently, the SLN radius $a$ was empirically obtained from $D_Z$ using the formula:
\begin{equation}
a \approx \frac{D_Z-\qty{91}{nm}}{2}
\end{equation}
This formula was assumed to be valid even for particles with $D_Z>\qty{200}{nm}$.

An estimate of $\Delta a$ for all SLN samples was obtained from the PdI values using Equation~(\ref{deltaD_z}):
\begin{equation}
\Delta a \sim \sqrt{PdI} \cdot a
\end{equation}

\begin{table}[h]
\caption{\label{tab:table3}
Mean hydrodynamic diameter ($D_Z$) and mean diameter measured ($D_C$) in cryo-EM images on $N_C$ SLNs for 7 samples.}
\begin{ruledtabular}
\begin{tabular}{lccc}
Dye-$R_{BY}$ & $D_Z \pm \Delta D_Z~(nm)$ & $D_C \pm \Delta D_C~(nm)$& $N_C$ \\
\hline
BY1-8\%& 128 $\pm$ 55 & 38 $\pm$ 19 & 300 \\
BY1-18\%& 134 $\pm$ 56 & 45 $\pm$ 23 & 251 \\
BY2-7\%& 125 $\pm$ 56 & 34 $\pm$ 19 & 167 \\
BY7-6\%& 121 $\pm$ 50 & 41 $\pm$ 22 & 160 \\
BY7-14\%& 128 $\pm$ 52 & 41 $\pm$ 26 & 396 \\
BY7-10\%& 199 $\pm$ 59 & 122 $\pm$ 75 & 82 \\
BY7-10\%& 200 $\pm$ 65 & 121 $\pm$ 80 & 114 \\
\end{tabular}
\end{ruledtabular}
\end{table}

\subsubsection{Particle number density.}

The particle number densities were measured with Nanoparticle tracking analysis (NTA) at \qty{25.0}{\degreeCelsius} (NanoSight LM10 \qty{405}{\nm}, Amesbury, UK) for one SLN batch. These measurements were performed to obtain the order of magnitude of the particle number density of the suspension used in the PA spectrometer and to verify the model applicability (section~\ref{sec:applic}). 

The particle number concentration of the mother solutions was evaluated $\sim 3 \times 10^{19}~\textrm{particles}.\unit{m^{-3}}$ for SLNs with a radius $a \sim \qty{35}{\nm}$ regardless of the BY-Palm label and $R_{BY}$, and $\sim 3 \times 10^{18}~\textrm{particles}.\unit{m^{-3}}$ for $a \approx \qty{122}{\nm}$ for SLNs BY1-10\%. These orders of magnitude are consistent with previously reported concentrations~\cite{linger_modulation_2025}. 
Therefore, for optically characterized diluted solutions $n_{sph} \sim 10^{17}-10^{18}~\textrm{particles}.\unit{m^{-3}}$. The mean inter-particular distance is given by $d_{inter} \sim {n_{sph}}^{-1/3} \sim \qty{1}{\um}$.

\section{\label{sec:applic}Application of the models}

This section gives the physical constants used for the validation of both the PA model (section~\ref{sec:PA_mod}) and the determination of the heat confinement ratio (section~\ref{sec:chi_mod}). Additionally, the conditions for the PA model (section \ref{sec:Mod3}) are verified.  

\subsection{Physical constants.}

\begin{table}[t]
\caption{\label{tab:table1}
Physical constants of water at three equilibrium temperature~\cite{haynes_crc_2017}.}
\begin{ruledtabular}
\begin{tabular}{lccc}
$T_{eq}~(\unit{\degreeCelsius})$& 15 & 25 & 37\\
\hline
$\Lambda_{wat}~(\unit{W.K^{-1}.m^{-1}})$& 0.589 & 0.606 & 0.625\\
$\kappa_{wat}~(\times 10^{-7}\unit{m^{2}.s^{-1}}$)& 1.41 & 1.46 & 1.50 \\
$\rho_{wat}~(\unit{kg. m^{-3}})$& 999 & 997 & 993 \\
$\beta_{wat}~(\times 10^{-4}\unit{K^{-1}})$& 1.51 & 2.57 & 3.62 \\
$C_{p,wat}~(\times 10^{3}\unit{J. kg^{-1}. K^{-1}})$& 4.19 & 4.18 & 4.18 \\
$c_{wat}~(\unit{m.s^{-1}})$& 1466 & 1497 & 1524 \\
\hline
$\xi_{wat}~(\times 10^{-11}\unit{Pa^{-1}})$\footnotemark[1]& 3.6 & 6.2 & 8.7 \\
$\Gamma_{wat}\footnotemark[1]$& 0.08 & 0.14 & 0.20 \\
\end{tabular}
\end{ruledtabular}
\footnotetext[1]{Computed with the other constants.}
\end{table}

\begin{table}[h]
\caption{\label{tab:table2}
Physical constants of solid palmitic acid at room temperature.}
\begin{ruledtabular}
\begin{tabular}{lc}
$\Lambda_{palm~acid}~(\unit{W. K^{-1}. m^{-1}})$\footnotemark[1]& 0.25 $\pm$ 0.01\\
$\kappa_{palm~acid}~(\times 10^{-7} \unit{m^{2}. s^{-1}}$)\footnotemark[2]& 1.4 \\
$\rho_{palm~acid}~(\unit{kg.m^{-3}})$\footnotemark[1]& 990 \\
$\beta_{palm~acid}~(\times 10^{-4} \unit{K^{-1}})$\footnotemark[3]& 3.1 \\
$C_{p,palm~acid}~(\times 10^{3} \unit{J.kg^{-1}. K^{-1}})$\footnotemark[1]& 1.8 \\
\hline
$\xi_{palm~acid}~(\times 10^{-11} \unit{Pa^{-1}})$\footnotemark[2]& 17.4 \\
\end{tabular}
\end{ruledtabular}
\footnotetext[1]{Reference~\cite{ma14164707}.}
\footnotetext[2]{Computed with the other constants.}
\footnotetext[3]{Reference~\cite{doi:10.1021/j150498a008}.}
\end{table}

Physical constants for water are tabulated~\cite{haynes_crc_2017} and Table~\ref{tab:table1} gathers the values at the three equilibrium temperatures set for the experimental PA measurements.

For the SLNs, the physical constants of the lipid core are not known as the physical properties of dexamethasone palmitate (DXP) are poorly documented in the literature and those of labelling BY-Palm agents were not experimentally determined.
However, both DXP and BY-Palm comprise a palmitate chain and are solid at the considered temperatures. The lipid core nanostructure of labelled DXP-SLN was reported to be mostly amorphous, with a short correlation length linked to the palmitate chain~\cite{linger_modulation_2025,sarah_preprint_2026}.

To obtain realistic values of the physical constants, we used an ``analog'' compound, whose thermodynamic properties are reported in the literature. Palmitic acid ($palm~acid$) was chosen because of its palmitate chain, its solid state at room temperature and because physical constants were reported by others~\cite{ma14164707,doi:10.1021/j150498a008}. DXP and palmitic acid have similar melting temperature as ``an endothermic peak between $49\unit{\degreeCelsius}$ and $68\unit{\degreeCelsius}$'' corresponding to the melting process was reported for DXP~\cite{doi_physicochemical_1989} and the melting point for palmitic acid is $62.5\unit{\degreeCelsius}$~\cite{haynes_crc_2017}. 
Table~\ref{tab:table2} gather the physical constants values for palmitic acid, which are used in the models for the SLNs.
The speed of sound in solid palmitic acid was not found in the literature. However, from phase velocity measurement in coconut oil~\cite{mori_monitoring_2020}, the longitudinal wave velocity in the lipid core of the SLNs can be estimated to: $c_{SLN} \gtrsim c_{wat}$.

\subsection{Application to the models}

\subsubsection{PA model.}
\emph{Condition on the particle size}:
  With estimated radius $a \le \qty{250}{\nm}$, $a/\tau_L \lesssim$ 40 \unit{m.s^{-1}}, which is much smaller than $c_{SLN}$ and $c_{wat}$. Then, the SLN samples are excited in long pulse regime.

\smallskip
\emph{Conditions on the particle concentration}:
\begin{itemize}
\item  The thermal diffusion length during $\tau_L$ in a spherical geometry is given by $\ell_{th}\sim \sqrt{6 \cdot \kappa \cdot \tau_L}$ with $\kappa$ the thermal diffusivity~\cite{mckenzie_physics_1990}. Then, $\ell_{th} \sim \qty{70}{\nm}$ both in the SLN core and in water. The spatial scale of the heated region around each SLN is then $(a + \ell_{th}) \lesssim \qty{320}{\nm}$, which is smaller than $d_{inter}$. Thereby, the heated regions do not overlap.
\item The volume fraction of the SLNs can be estimated by computing: $\frac{4}{3}\pi a^3 \cdot n_{sph}$ and is estimated here below 1\%.
\item $c_{wat} \cdot \tau_L \sim \qty{9}{\um}$ is greater than $d_{inter}$. Therefore, the acoustic field is continuous for the highest generated ultrasound frequencies. Regarding the highest detected ultrasound frequencies, the spatial scale is $\sim \qty{250}{\um}$, which is much greater than $d_{inter}$. Thereby, the acoustic field can be considered continuous in the sample.
 \end{itemize}

\subsubsection{Model of heat confinement ratio.}

Given the values of Tables~\ref{tab:table1} and~\ref{tab:table2}, no significant difference was found between the $\chi_{SLN}$ values computed using the physical constants of water at the three equilibrium temperature. Therefore, the $\chi_{SLN}$ values evaluated at $T_{eq}= 25\unit{\degreeCelsius}$ were used to fit the estimated $PGE$ values.

\bibliography{size260707}